\documentclass[preprint,proceedings]{rmaa}

\usepackage{paralist}

\usepackage{psfrag,color}

\newcommand{\ie}{\emph{i.e.}\ }
\newcommand{\eg}{\emph{e.g.}\ }

\newcommand{\msun}{M$_\odot$}

\SetYear{2002}
\SetConfTitle{Galaxy Evolution: Theory and Observations}

\title{Observational Properties of Primordial Stellar Populations}
\author{
  N. Panagia,\altaffilmark{1,2} 
  M. Stiavelli,\altaffilmark{1}
  H. Ferguson,\altaffilmark{1}
  and H.S. Stockman\altaffilmark{1}}

\altaffiltext{1}{Space Telescope Science Institute, Baltimore, MD 21218,
USA.
}
\altaffiltext{2}{On assignment from the Research and Scientific Support
Department of ESA.
}

\shortauthor{Panagia et al.}
\shorttitle{Primordial Stellar Populations}

\fulladdresses{
\item H.C. Ferguson, N. Panagia, M. Stiavelli, \& H.S. Stockman, Space
Telescope Science Institute, 3700 San Martin Drive, Baltimore, MD 21218,
USA (ferguson, panagia, mstiavel, stockman@stsci.edu).}

\listofauthors{N. Panagia, M. Stiavelli, H. Ferguson, \& H.S. Stockman}
\indexauthor{Panagia, N.}
\indexauthor{Stiavelli, M.}
\indexauthor{Ferguson, H.C.}
\indexauthor{Stockman, H.S.}

\abstract{We present the first results of a study of the expected properties  of
the first stellar generations in the Universe. In particular, we
consider and discuss a series of properties that, on the basis of the
emission from associated HII regions,  permit one to discern {\it bona
fide} primeval stellar generations from the ones formed after pollution
from supernova explosions. The expected performance of NGST for the
study and the characterization of primordial sources is also discussed.}

\addkeyword{Cosmology: early universe}
\addkeyword{Cosmology: observations}
\addkeyword{Galaxies: abundance}
\addkeyword{Galaxies: starburst}
\addkeyword{H~II regions}
\begin{document}
\maketitle

\section{Primordial Stars: Expected Properties}
\label{sec:PrimStars}

The standard picture is that at zero metallicity the Jeans mass in star
forming clouds is much higher than it is in the local Universe, and,
therefore, the formation of massive stars, say, 100 \msun\/ or higher,
is highly favored. The spectral distributions (SED) of these massive
stars are characterized by effective temperatures on the Main Sequence
(MS) around $10^5$~K (Tumlinson \& Shull 2000, Bromm \etal~2001, Marigo
\etal~2001).  Due to their temperatures these stars are very effective
in ionizing hydrogen and helium. It should be noted that
zero-metallicity (the so-called population III) stars  of all masses
have essentially the same MS luminosities as, but are much hotter than
their solar metallicity analogues.   Note also that only stars hotter
than about 90,000~K are capable of ionizing He twice in appreciable
quantities, say, more than about 10\% of the total He content (\eg
Oliva \& Panagia 1983, Tumlinson \& Shull 2000).  As a consequence even
the most massive population III stars can produce HeII lines only for a
relatively small fraction of their lifetimes, say, about 1~Myr or about
1/3 of their lifetimes.

The second generation of stars forming out of pre-enriched material
will probably have different properties because cooling by metal lines
may become a viable mechanism and  stars of lower masses may be
produced (Bromm \etal\/ 2001). On the other hand, if the metallicity is
lower than about $5\times 10^{-4}$Z$_\odot$,  build up of H$_2$ due to
self-shielding may occur, thus  continuing the formation of very
massive stars (Oh \& Haiman 2002). Thus, it appears that in the
zero-metallicity case one may  expect a very top-heavy Initial Mass
Function (IMF), whereas it is not clear if the second generation of stars
is also top-heavy or follows a normal IMF. 


\section{Primordial HII Regions}

The  high effective temperatures of zero-metallicity stars imply not
only high ionizing photon fluxes for both hydrogen and helium, but
also  low optical  and UV fluxes.  This is because the optical/UV
domains fall  in the Rayleigh-Jeans tail  of the spectrum where the
flux is proportional to the first power of the  effective temperature,
$T_{eff}$,  so that, for equal bolometric luminosity, the actual flux
scales like $T_{eff}^{-3}$.  Therefore, an average increase of
effective temperature of a factor of $\sim2$ will give a reduction of
the optical/UV flux by a factor of $\sim8$. As a result,  one should
expect the rest-frame optical/UV spectrum of a primordial HII regions
to be largely dominated by its nebular emission (both continuum and
lines), so that the best strategy to detect the presence of primordial
stars is to search for the emission from associated HII regions. 

  \begin{figure}
    \begin{center}
      \includegraphics[width=\columnwidth]{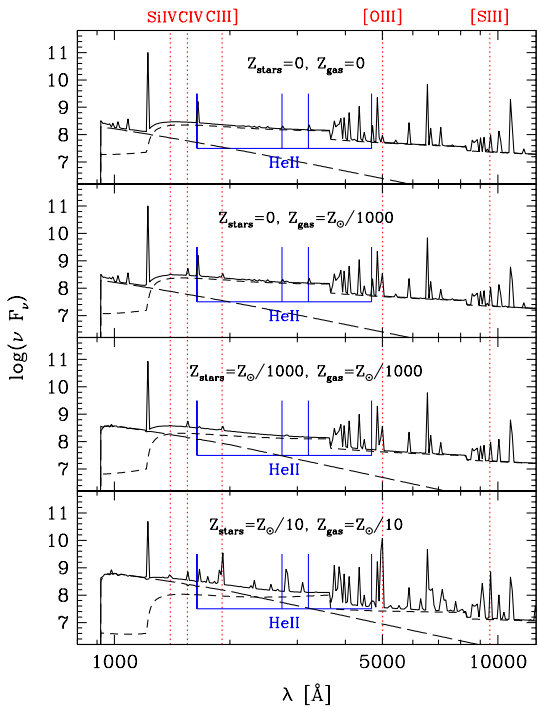}
    \end{center}
  \caption{{The synthetic spectrum of a zero-metallicity HII region
(top panel) is compared to that of HII regions with  various
combinations of stellar and nebular metallicities (lower panels). The
long-dashed and short-dashed lines represent the stellar and nebular
continua, respectively. } \label{{label}} } 
  \end{figure}
%
%

In Panagia \etal\/(2002) we report on our calculations using Cloudy90
(Ferland \etal\/1998) of the properties of primordial, zero-metallicity
HII regions (\eg Figure~1). We find that the electron temperatures is
in excess of 20,000 K and that 45\% of the total luminosity is converted
into the Ly-$\alpha$ line, resulting in a Ly-$\alpha$ equivalent width
(EW) of 3000 \AA\/  (Bromm, Kudritzki \& Loeb 2001). The helium lines are
also strong, with the HeII $\lambda$1640 intensity comparable to
that of H$\beta$ (Panagia \etal\/2002, Tumlinson \etal\/2001).

An interesting feature of these models is that the continuum longwards
of Ly-$\alpha$ is dominated by the two-photon nebular continuum.
The H$\alpha$/H$\beta$ ratio for these models is 3.2. Both the red
continuum and the high H$\alpha$/H$\beta$ ratio could be naively (and
incorrectly) interpreted as a consequence of dust extinction even
though no dust is present in these systems.

From the observational point of view one will generally be unable to
measure a zero-metallicity but will usually be able to place an upper
limit to it. When would such an upper limit be indicative that one is
dealing with a population III object? According to Miralda-Escud\'e \&
Rees (1998) a metallicity Z$\simeq10^{-3}Z_\odot$ can be used as a
dividing line between the pre- and post-re-ionization Universe. A
similar value is obtained by considering that the first supernova (SN)
going off in a primordial cloud will pollute it to a metallicity of
$\sim 0.5 \times 10^{-3}Z_\odot$\ (Panagia \etal\/2002). Thus, any object
with a metallicity higher than $\sim 10^{-3} Z_\odot$ is not a true
first generation object.


  \begin{figure}
    \begin{center}
      \includegraphics[width=\columnwidth]{{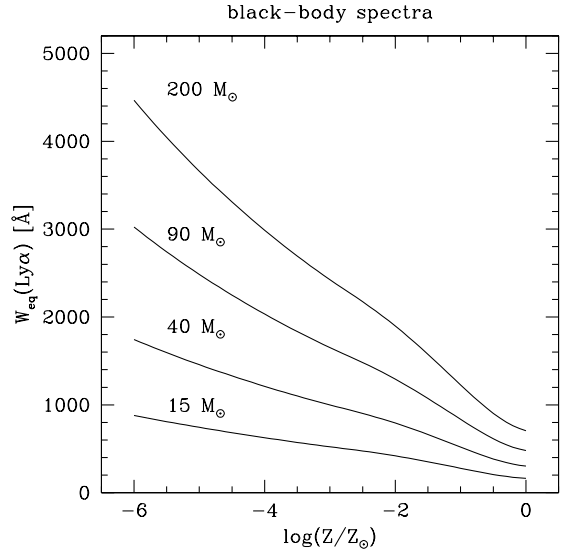}}
    \end{center}
  \caption{{Ly-$\alpha$ equivalent widths for HII regions ionized by stars with
a range of masses and metallicities. The results obtained for black bodies
or stellar atmospheres are very similar.} \label{{label}} } 
  \end{figure}
   


  \begin{figure}
    \begin{center}
       \includegraphics[width=\columnwidth]{{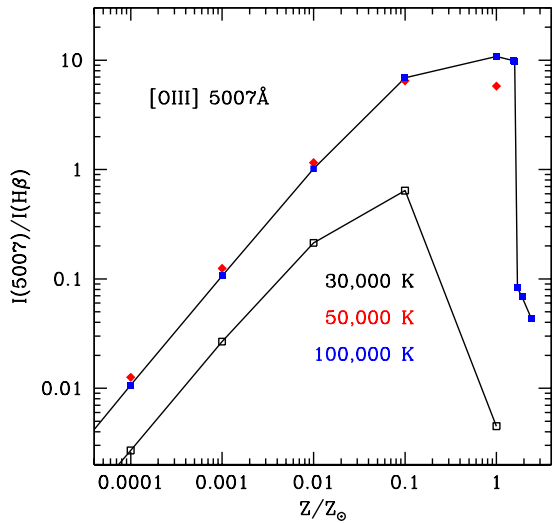}}
   \end{center}
  \caption{{The ratio [OIII]$\lambda 5007$ / H$\beta$ is plotted as a function of
metallicity for three different stellar masses: 30,000K (open squares and
bottom line), 50,000K (solid diamonds), and 100,000K (solid squares and top
line).} \label{{label}} } 
  \end{figure}

\section{Low Metallicity HII Regions}

 We have also computed model HII regions for metallicities from three
times solar  down to $10^{-6} Z_\odot$\ (Panagia \etal\/ 2002).  In
Figure~1 the synthetic spectrum of an HII region with metallicity
$10^{-3} Z_\odot$ (third panel from the top) can be compared to that
of  an object with zero metallicity (top panel). The two are very
similar except for a few weak metal lines.  In Figure~2 we show the
Ly-$\alpha$ EWs for HII regions ionized by stars with a range of
stellar masses and metallicities. Values of EW in excess of 1,000\AA\/
are possible already for objects with metallicity $\sim 10^{-3}
Z_\odot$. This is particularly interesting given that Ly-$\alpha$
emitters with large EW have been identified at z=5.6 (Rhoads \&
Malhotra 2001).

The metal lines are weak, but some of them can be used as metallicity
tracers. In Figure~3 the intensity ratio of the  [OIII]$\lambda 5007$
line to H$\beta$ is plotted for a range of stellar temperatures and
metallicities. It is apparent that for $Z < 10^{-2} Z_\odot$ this line
ratio traces metallicity linearly. Our reference value $Z = 10^{-3}$
corresponds to a ratio [OIII]/H$\beta$ = 0.1. The weak dependence on
stellar temperature makes sure that this ratio remains a good indicator
of metallicity also when one considers the integrated signal from a
population with a range of stellar masses.


  \begin{figure}
    \begin{center}
      \includegraphics[width=\columnwidth]{{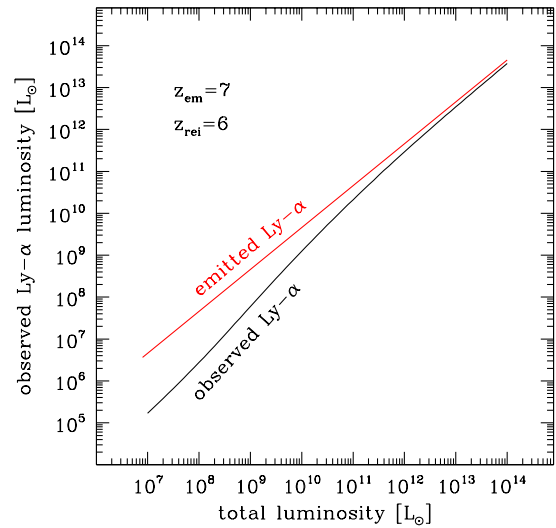}}
    \end{center}
  \caption{{Transmitted Ly-$\alpha$ intensity as a function of the object
luminosity. Bright objects ionize their neighborhood and are
able to reduce the Ly-$\alpha$ attenuation.} \label{{label}} } 
  \end{figure}
 

Another difference between zero-metallicity and low-metallicity HII
regions lies in the possibility that the latter may contain dust. For
a $Z=10^{-3} Z_\odot$ HII region dust may absorb up to 30 \% of the
Ly-$\alpha$ line, resulting in roughly 15 \% of the energy being
emitted in the far IR (Panagia \etal~2002).

\section{How to discover and characterize  Primordial HII Regions}

It is natural to wonder whether primordial HII regions will be
observable with the generation of telescopes currently on the drawing
boards. In this section we will focus mostly on the capabilities of the
Next Generation Space Telescope.

Before proceeding further we have to include he effect of HI absorption
in the IGM on the Ly-$\alpha$ radiation (Miralda-Escud\'e \& Rees 1998,
Madau \& Rees 2001, Panagia \etal\/2002).  A comparison of the observed
vs emitted Ly-$\alpha$ intensities is given in Figure~4.  The
transmitted Ly-$\alpha$ flux depends on the total luminosity of the
source since this determines the radius of the resulting Str\"omgren
sphere. A Ly-$\alpha$  luminosity of $\sim10^{10}$ L$_\odot$
corresponds to $\sim10^6$ M$_\odot$ in massive stars. In the following
we will consider this as our reference model.

The synthetic spectra, convolved with suitable filter responses  can be
compared directly to the NGST imaging sensitivity for $4\times10^5$s
exposures (see Figure~5). It is clear that NGST will be able to easily
detect such objects. Due to the high background from the ground, NGST
will remain superior even to 30m ground based telescopes for these
applications.


  \begin{figure}
    \begin{center}

 \includegraphics[width=\columnwidth]{{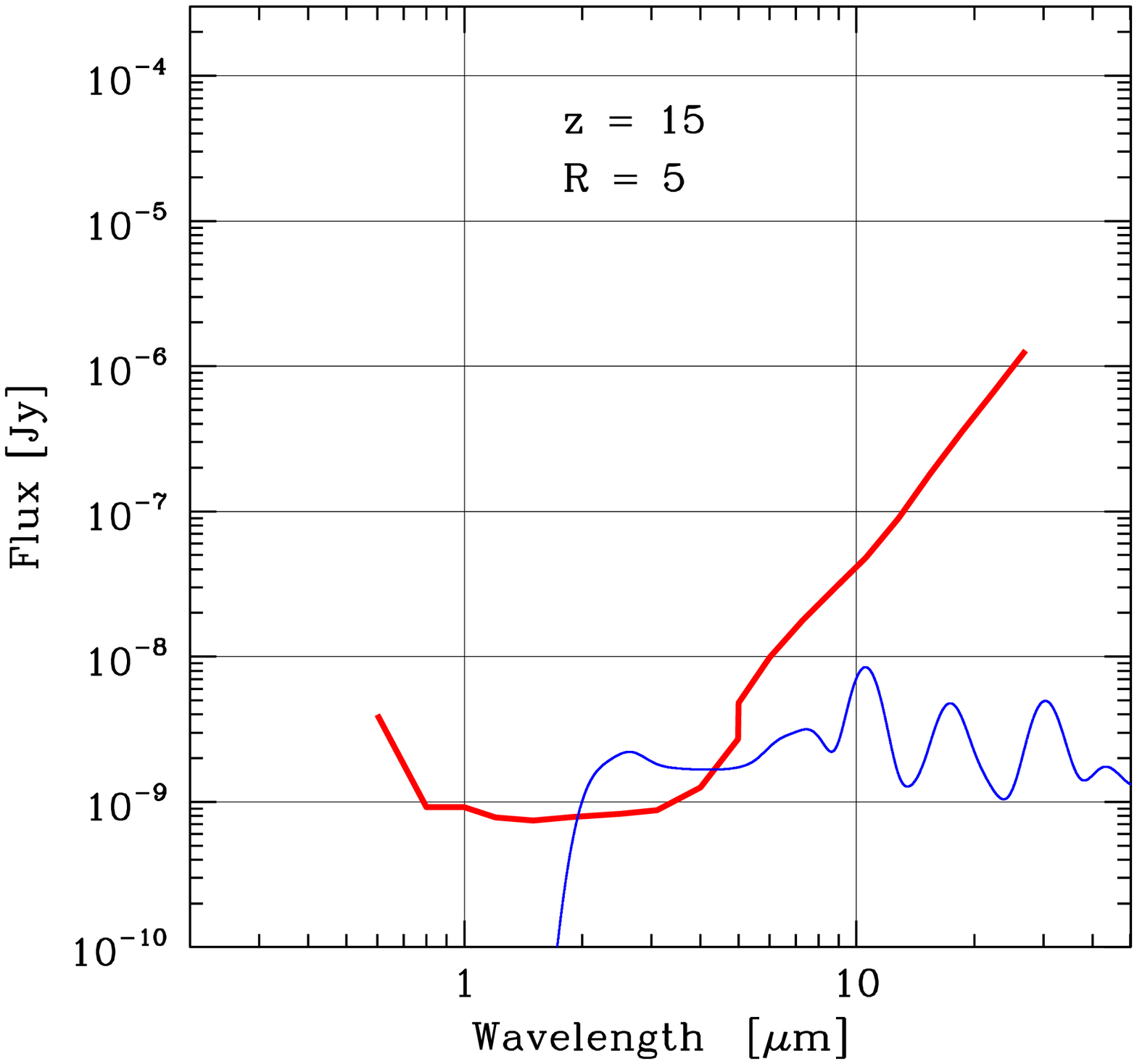}}
   \end{center}
  \caption{{Synthetic spectral energy distribution of a Z=$10^{-3} Z_\odot$
starburst object at z=15 containing $10^6$ M$_\odot$ in massive stars
(thin line) compared to the imaging limit of NGST at R=5 (thick line).
The NGST sensitivity refers to $4\times10^5$ s exposures with S/N=10t.} \label{{label}} } 
  \end{figure}


  \begin{figure}
    \begin{center}

\includegraphics[width=\columnwidth]{{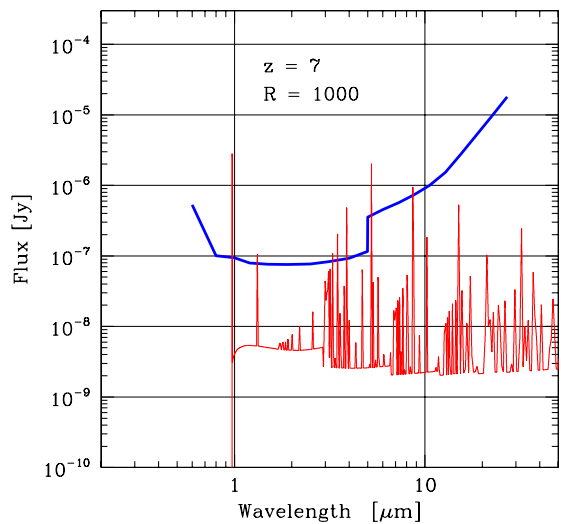}}
    \end{center}
  \caption{{Synthetic spectrum of a Z=$10^{-3} Z_\odot$ starburst object at z=7
containing $10^6$ M$_\odot$ in massive stars (thin line) compared to the
spectroscopic limit of NGST at R=1000 (thick line). The NGST sensitivity
refers to $4\times10^5$ s exposures with S/N=10.} \label{{label}} } 
  \end{figure}


%
  \begin{figure}
    \begin{center}
      \includegraphics[width=\columnwidth]{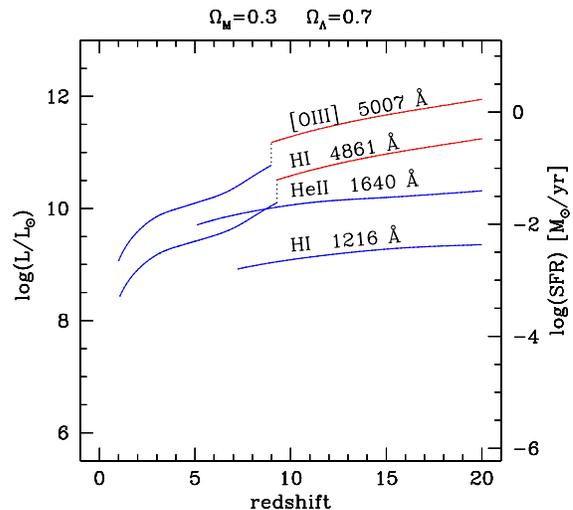}
    \end{center}
  \caption{{Limiting {\it total} luminosity of the ionizing stars (left-hand scale) and
  top-heavy IMF star formation rate (right-hand scale) to detect various
  emission lines using NGST spectroscopy, with S/N=10 in integrations of
  100 hours, as a function of the source redshift.} \label{{label}} } 
  \end{figure}
%
%

The synthetic spectra can also be compared to the NGST spectroscopic
sensitivity for $4\times10^5$s exposures (see Figure 6): it appears
that while the  Ly-$\alpha$ line can be detected up to redshifts as
high as 15 or 20, for our reference source only at relatively low
redshifts (z$\sim7$), can NGST detect other diagnostics lines lines
such as HeII 1640\AA, and Balmer lines. Determining metallicities is
then limited to either lower redshifts or to brighter sources. 

We can reverse the argument and ask ourselves what kind of sources can
NGST detect and characterize with spectroscopic observations.  Figure~7
displays, as a function of redshift,  the total luminosity of a
starburst  whose lines can be detected with a S/N=10 adopting an
exposure time of $4\times10^5$s. The loci for Ly-$\alpha$, HeII
1640\AA, H\/$\beta$, and [OIII] 5007\AA\/ are shown. It appears that
Ly-$\alpha$ is readily detectable up to z$\simeq$20, HeII 1640\AA\/ may
also be detected up to high redshifts {\it if} massive stars are indeed
as hot as predicted, whereas ``metallicity" information, \ie the
intensity ratio I([OIII])/I(H$\beta$), can be obtained at high
redshifts only for sources that are 10--100 times more massive or that
are 10--100 times  magnified by gravitational lensing.  

\section{Conclusions}
We have considered and discussed a series of properties that, on the
basis of the emission from associated HII regions,  permit one to
discern {\it bona fide} primeval stellar generations from the ones
formed after pollution from supernova explosions. We find that it is
possible to discern truly primordial populations from the next
generation of stars by measuring the metallicity of high-z star forming
objects. The very low background of NGST will enable it to image and
study first-light sources at very high redshifts, whereas its
relatively small collecting area olimits its capability in obtaining
spectra of z$\sim$10--15 first-light sources to either the bright end
of their luminosity function or to strongly lensed sources.

\end{document}